\DeclareUnicodeCharacter{2212}{-}
\RequirePackage{luatex85}

\documentclass[twocolumn]{aastex63}

\usepackage[caption=false]{subfig}
\usepackage{tabularx}
\usepackage{comment}
\usepackage{pifont}
\usepackage{natbib}
\usepackage[autostyle]{csquotes}
\usepackage{CJKutf8}

\submitjournal{ApJ}
\graphicspath{{./}{figures/}}

\begin{document}

\title{SN~2022joj: A Potential Double Detonation with a Thin Helium shell}
\shorttitle{SN~2022joj: A Peculiar Type Ia}
\shortauthors{Padilla et al.}

\correspondingauthor{Estefania Padilla Gonzalez}
\email{epadillagonzalez@ucsb.edu}

\author[0000-0003-0209-9246]{E. Padilla Gonzalez}
\affiliation{Department of Physics, University of California, Santa Barbara, CA 93106-9530, USA}
\affiliation{Las Cumbres Observatory, 6740 Cortona Dr, Suite 102, Goleta, CA 93117-5575, USA}

\author[0000-0003-4253-656X]{D.A. Howell}
\affiliation{Department of Physics, University of California, Santa Barbara, CA 93106-9530, USA}
\affiliation{Las Cumbres Observatory, 6740 Cortona Dr, Suite 102, Goleta, CA 93117-5575, USA}

\author[0000-0003-0794-5982]{G. Terreran}
\affiliation{Las Cumbres Observatory, 6740 Cortona Dr, Suite 102, Goleta, CA 93117-5575, USA}

\author[0000-0001-5807-7893]{C. McCully}
\affiliation{Department of Physics, University of California, Santa Barbara, CA 93106-9530, USA}
\affiliation{Las Cumbres Observatory, 6740 Cortona Dr, Suite 102, Goleta, CA 93117-5575, USA}

\author[0000-0001-9570-0584]{M. Newsome}
\affiliation{Department of Physics, University of California, Santa Barbara, CA 93106-9530, USA}
\affiliation{Las Cumbres Observatory, 6740 Cortona Dr, Suite 102, Goleta, CA 93117-5575, USA}

\author[0000-0003-0035-6659]{J. Burke}
\affiliation{Department of Physics, University of California, Santa Barbara, CA 93106-9530, USA}
\affiliation{Las Cumbres Observatory, 6740 Cortona Dr, Suite 102, Goleta, CA 93117-5575, USA}

\author[0000-0003-4914-5625]{J. Farah}
\affiliation{Department of Physics, University of California, Santa Barbara, CA 93106-9530, USA}
\affiliation{Las Cumbres Observatory, 6740 Cortona Dr, Suite 102, Goleta, CA 93117-5575, USA}

\author[0000-0002-7472-1279]{C. Pellegrino}
\affiliation{Department of Physics, University of California, Santa Barbara, CA 93106-9530, USA}
\affiliation{Las Cumbres Observatory, 6740 Cortona Dr, Suite 102, Goleta, CA 93117-5575, USA}

\newcommand{\UA}{\affiliation{Steward Observatory, University of Arizona, 933 North Cherry Avenue, Tucson, AZ 85721-0065, USA}}
\newcommand{\Catalyst}{\altaffiliation{LSSTC Catalyst Fellow}}
\author[0000-0002-4924-444X]{K. A. Bostroem}
\Catalyst\UA
\author[0000-0002-0832-2974]{G. Hosseinzadeh}
\UA
\author[0000-0002-0744-0047]{J. Pearson}
\UA
\author[0000-0003-4102-380X]{D. J. Sand}
\UA
\author[0000-0002-4022-1874]{M. Shrestha}
\UA
\author[0000-0001-5510-2424]{N. Smith}
\UA

\author[0000-0002-7937-6371]{Y. Dong \begin{CJK*}{UTF8}{gbsn}(董一泽)\end{CJK*}}
\affiliation{Department of Physics and Astronomy, University of California, 1 Shields Avenue, Davis, CA 95616-5270, USA}

\author[0000-0002-7015-3446]{N. Meza Retamal}
\affiliation{Department of Physics and Astronomy, University of California, 1 Shields Avenue, Davis, CA 95616-5270, USA}

\author[0000-0001-8818-0795]{S. Valenti}
\affiliation{Department of Physics and Astronomy, University of California, 1 Shields Avenue, Davis, CA 95616-5270, USA}

\author[0000-0002-1184-0692]{S. Boos}
\affiliation{Department of Physics and Astronomy, University of Alabama, Tuscaloosa, AL, 35487, USA}

\author[0000-0002-9632-6106]{K. J. Shen}
\affiliation{Department of Astronomy and Theoretical Astrophysics Center, University of California, Berkeley, CA 94720, USA}

\author[0000-0002-9538-5948]{D. Townsley}
\affiliation{Department of Physics and Astronomy, University of Alabama, Tuscaloosa, AL, 35487, USA}

\author[0000-0002-1296-6887]{L. Galbany}
\affiliation{Institute of Space Sciences (ICE-CSIC), Campus UAB, Carrer de Can Magrans, s/n, E-08193 Barcelona, Spain.}
\affiliation{Institut d’Estudis Espacials de Catalunya (IEEC), E-08034 Barcelona, Spain.}

\author[0000-0002-1296-6887]{L. Piscarreta}
\affiliation{Institute of Space Sciences (ICE-CSIC), Campus UAB, Carrer de Can Magrans, s/n, E-08193 Barcelona, Spain.}

\author[0000-0002-2445-5275]{R.~J.~Foley}
\affiliation{Department of Astronomy and Astrophysics, University of California, Santa Cruz, CA 95064, USA}

\author[0000-0003-0416-9818]{M.~J.~Bustamante-Rosell}
\affiliation{Department of Astronomy and Astrophysics, University of California, Santa Cruz, CA 95064, USA}

\author[0000-0003-4263-2228]{D.~A.~Coulter}
\affiliation{Department of Astronomy and Astrophysics, University of California, Santa Cruz, CA 95064, USA}

\author[0000-0002-7706-5668]{R. ~Chornock}
\affiliation{Department of Astronomy, University of California, Berkeley, CA 94720-3411, USA}

\author[0000-0002-5680-4660]{K.~W.~Davis}
\affiliation{Department of Astronomy and Astrophysics, University of California, Santa Cruz, CA 95064, USA}

\author[0000-0001-9749-4200]{C.~B.~Dickinson}
\affiliation{Department of Astronomy and Astrophysics, University of California, Santa Cruz, CA 95064, USA}

\author[0000-0002-6230-0151]{D.~O.~Jones}
\affiliation{Gemini Observatory/NSF’s NOIRLab, 670 N. A’ohoku Place, Hilo, HI
96720, USA}
\author[0009-0004-7605-8484]{J.~Kutcka}
\affiliation{Department of Astronomy and Astrophysics, University of California, Santa Cruz, CA 95064, USA}

\author[0009-0004-3242-282X]{X.K. Le Saux}
\affiliation{Department of Astronomy and Astrophysics, University of California, Santa Cruz, CA 95064, USA}

\author[0000-0002-7559-315X]{C.~R.~Rojas-Bravo}
\affiliation{Department of Astronomy and Astrophysics, University of California, Santa Cruz, CA 95064, USA}

\author[0000-0002-5748-4558]{K.~Taggart}
\affiliation{Department of Astronomy and Astrophysics, University of California, Santa Cruz, CA 95064, USA}

\author[0000-0002-1481-4676]{S. Tinyanont}
\affiliation{National Astronomical Research Institute of Thailand, 260  Moo 4, Donkaew,  Maerim, Chiang Mai, 50180, Thailand}

\author[0000-0001-7823-2627]{G.~Yang}
\affiliation{The Thacher School, 5025 Thacher Road, Ojai, CA 93023, USA}

\author[0000-0001-8738-6011]{S. W.~Jha}
\affiliation{Department of Physics and Astronomy, Rutgers, the State University of New Jersey, Piscataway NJ, USA}

\author[0000-0003-4768-7586]{R. Margutti}
\affiliation{Department of Astronomy, University of California, Berkeley, CA 94720-3411, USA}
\affiliation{Department of Physics, University of California, 366 Physics North MC 7300, Berkeley, CA 94720, USA}

\begin{abstract}

We present photometric and spectroscopic data for SN~2022joj, a nearby peculiar Type Ia supernova (SN Ia) with a fast decline rate ($\rm{\Delta m_{15,B}=1.4}$~mag). SN~2022joj shows exceedingly red colors, with a value of approximately ${B-V \approx 1.1}$~mag during its initial stages, beginning from $11$~days before maximum brightness. As it evolves the flux shifts towards the blue end of the spectrum, approaching ${B-V \approx 0}$~mag around maximum light. Furthermore, at maximum light and beyond, the photometry is consistent with that of typical SNe Ia. This unusual behavior extends to its spectral characteristics, which initially displayed a red spectrum and later evolved to exhibit greater consistency with typical SNe Ia. We consider two potential explanations for this behavior: double detonation from a helium shell on a sub-Chandrasekhar-mass white dwarf and Chandrasekhar-mass models with a shallow distribution of $\rm{^{56}Ni}$. The shallow nickel models could not reproduce the red colors in the early light curves. Spectroscopically, we find strong agreement between SN~2022joj and double-detonation models with white dwarf masses around 1 $\rm{M_{\odot}}$ and thin He-shell between 0.01 and 0.02 $\rm{M_{\odot}}$. Moreover, the early red colors are explained by line-blanketing absorption from iron-peak elements created by the double detonation scenario in similar mass ranges. However, the nebular spectra composition in SN~2022joj deviates from expectations for double detonation, as we observe strong [\ion{Fe}{3}] emission instead of [\ion{Ca}{2}] lines as anticipated from double detonation models. More detailed modeling, e.g., including viewing angle effects, is required to test if double detonation models can explain the nebular spectra.
\end{abstract}


\section{Introduction} \label{sec:intro}
Type Ia supernovae (SNe Ia) play a pivotal role in measuring cosmological distances and were instrumental in revealing the acceleration of the universe \citep{Riess1998, Perlmutter1999}. Furthermore, they are responsible for synthesizing the majority of the iron-group elements (titanium through zinc) found throughout the universe \citep{Iwamoto1999}. Although extensively researched, their progenitor systems continue to be elusive: the nature of the companion star (degenerate \citealt{Iben&Tutukov1984, Webbik1984} or not \citealt{Whelan&Ibel1973}) and the explosion mechanism (surface detonation \citealt{Bildsten07} or core ignited deflagration transitioning to a detonation \citealt{Khokhlov1991, Woosley1994}) are all subject to debate. 

Studies have explored various explosion mechanisms for WDs including a helium detonation on the surface of a sub-Chandrasekhar mass WD. In this scenario, a sub-Chandrasekhar mass WD accumulates He from a helium star (or He WD or C/O WD, which have helium surface layers). This accumulation has the potential to trigger an ignition at the base of the He shell.  This has been suggested to be the cause of some faint and fast evolving transients such as Ca-rich transients \citep{Perets10}. The surface explosion may also drive a shock into the core causing a subsequent detonation \citep[also known as double detonation;][]{Bildsten07, Fink10, sim12,Shen14}. Observables can vary significantly depending on the thickness of the He shell. 

One predicted signature from a surface detonation of a thick helium shell ($\gtrsim$ 0.05 $\rm{M_{\odot}}$) is the production of a significant amount of $\rm{^{56}Ni}$, resulting in extreme UV line blanketing \citep{Kromer2010, sim12, Polin2019}. However, it was also shown by \cite{Sim10, Shen18, Townsley2019,Boos2021, kenshen2021} that double detonation models approaching the limiting case of a bare CO WD detonation are capable of roughly reproducing observations of SNe Ia. This is attributed to thin shell detonations mostly producing intermediate mass elements, instead of $\rm{^{56}Ni}$, and thus have less of an effect on the observables from the underlying core ashes.

In this paper, we discuss the peculiar SNe Ia 2022joj. This object is unique due to the strong reddening in the early spectra and light curves pointing towards a double detonation candidate. This paper is arranged as follows. In \S \ref{sec:discovery}, we discuss the discovery and observations. In \S \ref{sec:Analysis}, we discuss reddening, light curve and spectral analysis. In \S \ref{sec:Modeling}, we compare the light curves and spectra with double detonation models and variations of $\rm{^{56}Ni}$ distributions. Finally, in \S \ref{sec:discussion}, we discuss the possible progenitor system and conclude in \S \ref{sec:conclusion}.

\section{Discovery and Observations}\label{sec:discovery}
\subsection{Discovery}\label{sec:discover}

\begin{figure}
\includegraphics[width=0.45\textwidth]{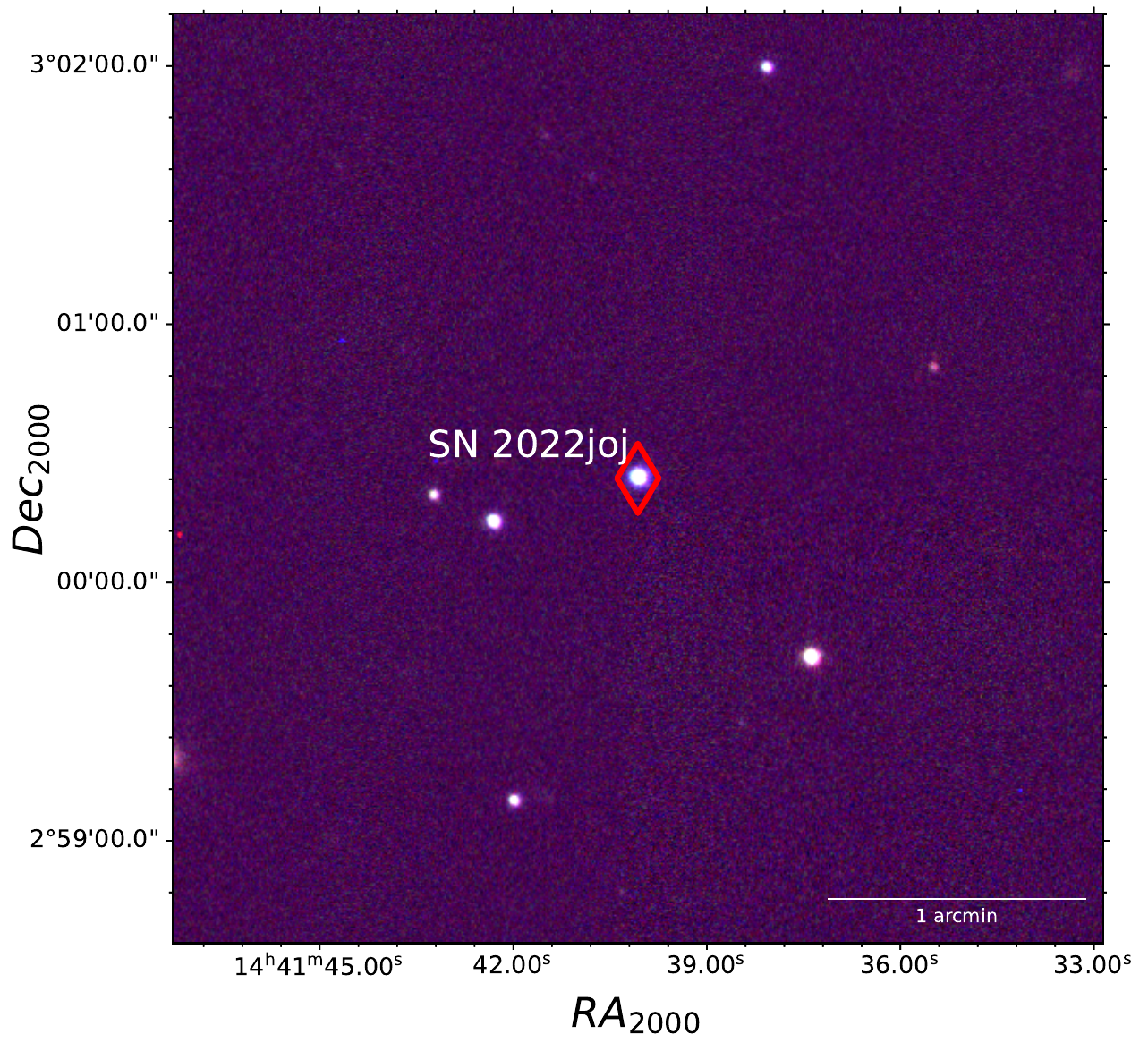}
\caption{$rVB$ stacked image of SN~2022joj taken on 2022 May 23. The image indicates the location of the supernova with a red diamond and --- the host galaxy is faint so is dominated by the supernova here.
\label{fig:image}}
\end{figure}

SN~2022joj was discovered by the Zwicky Transient Facility \citep[ZTF;][]{ZTF2021} on 2022 May 8.3 UT, using the ZTF survey at the Palomar Observatory with a discovery magnitude of 19.1 in the $r$ filter \citep{2022TNSTR1220....1F}. The last non-detection of the same object was on 2022 May 2.34 UT, with a $g$ magnitude of 20.25 \citep{2022TNSTR1220....1F}. An explosion time of 2459705.64 JD (or May 6, 2022) was estimated from a power-law fits with an index of 2 to the light curve in the $r$ band from time ranging between May 8 and May 14, 2022. SN~2022joj is located at right ascension $14^\mathrm{h}41^\mathrm{m}40\fs07$ and declination $+03^{\circ}00'24\farcs2$ shown in Figure \ref{fig:image} and discussed in more detail in Section \ref{sec:remote_location}. 

The Milky Way extinction value $E(B-V) = 0.0313$ mag was adopted from the \citet{Schlafly_2011} calibration of the \citet{Schlegel_1998} dust maps. \citet{2022TNSCR1274....1N} reported a classification on behalf of the Global Supernova Project  based on a spectrum acquired on May 11, 2022, at 11.28 UT, using FLOYDS on Las Cumbres Observatory's \citep[LCO;][]{Brown13} Faulkes Telescope North.  SN~2022joj was classified as a Type I SN due to its lack of hydrogen, despite its uncertainty due to its spectral peculiarity. These distinctive features include suppression in the blue between 3000 to 5000 $\rm{\AA}$ and the faint presence of O I. 


\subsection{Photometry}

\begin{figure}
\includegraphics[width=0.45\textwidth]{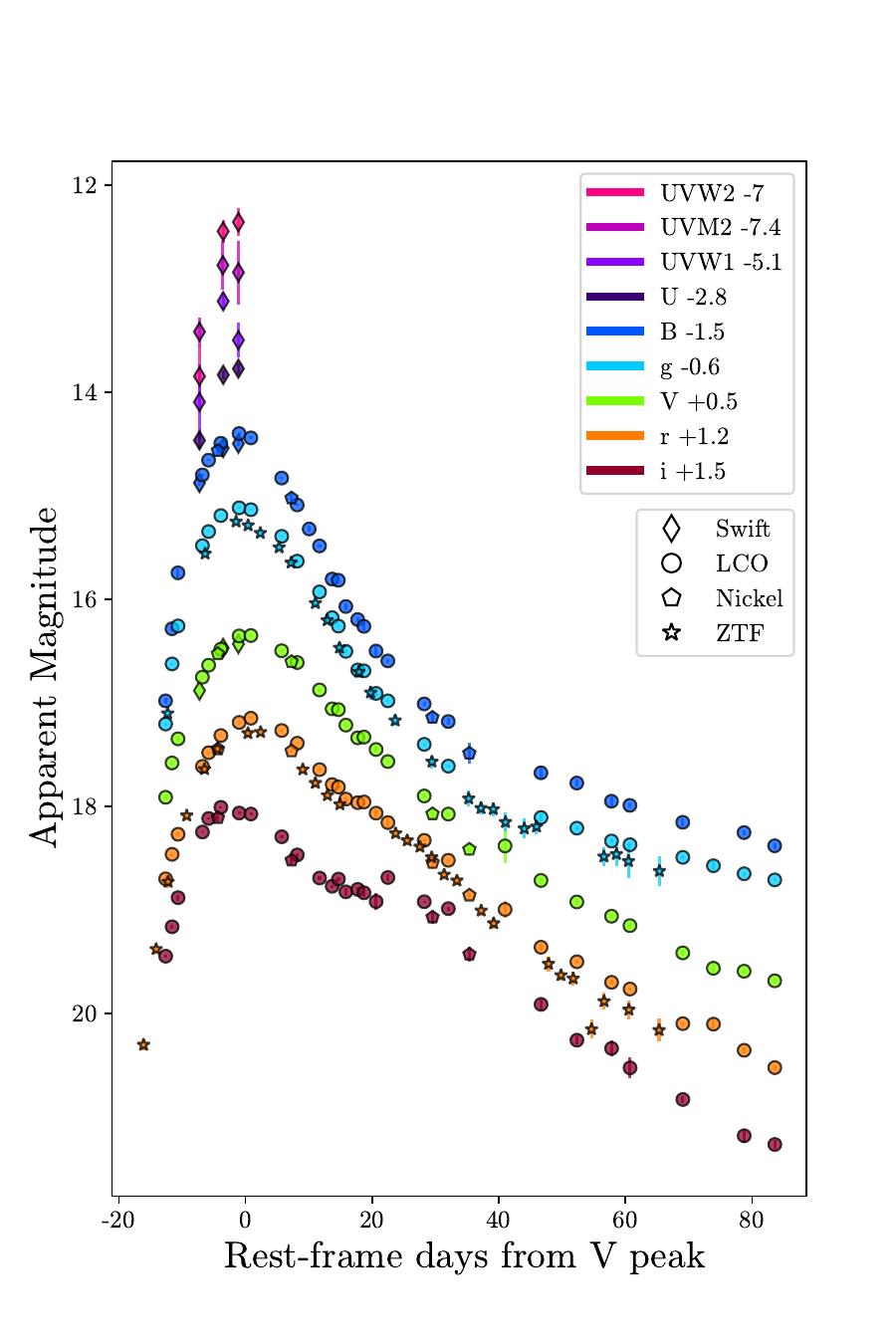}
\caption{Milky Way extinction corrected photometry using $E(B-V)_{\rm{MW}}=0.0313$ mag as discussed in Section \ref{sec:discovery}. These data are not corrected for host extinction (see Section \ref{sec:3.1} for details on host extinction).
\label{fig:22joj_lc}}
\end{figure}

Following the discovery of SN~2022joj, the Global Supernova Project triggered follow up observations and acquired photometry in the \textit{BVgri} filters using the 1-m telescopes from the LCO global network of telescopes. The data were reduced using \texttt{lcogtsnpipe} \citep{Valenti16}. The method utilized during the reduction was PSF-fitting photometry. The \textit{BV} zero points were calculated from the AAVSO Photometric All-Sky Survey (APASS) standard catalogue \citep{APASS2009}, whereas the \textit{gri} zeropoints were calculated from the Sloan magnitudes of field stars \citep{Albareti17}.
Additional optical photometry was retrieved from the publicly available ZTF survey\footnote{\url{https://alerce.online/object/ZTF22aajijjf}}.
In addition, imaging of SN~2022joj was obtained in $BVri$ bands with the 1~m Nickel telescope at Lick Observatory. The images were calibrated using bias and sky flat-field frames following standard procedures. PSF photometry was performed, and photometry was calibrated relative to Pan-STARRS photometric standards \citep{Flewelling16}. The full light curve is presented in Figure \ref{fig:22joj_lc}.
The $BV$ filters are calibrated to Vega \citep{Bessel1990}, whereas the $gri$ are calibrated in the AB magnitude system \citep{Fukigita1996}.

SN\,2022joj was observed with the Ultraviolet Optical Telescope (UVOT; \citealt{Roming05}) on board the Neil Gehrels Swift Observatory \citep{Gehrels04} for 4 epochs, from 2022 May 17.4 until 2022 May 22.7 ($\delta t= -5.4$--1.0\,days from \textit{B}-band maximum). We performed aperture photometry using a 5$\arcsec$-radius circular region with \texttt{uvotsource} within HEAsoft v6.26,\footnote{We used the calibration database (CALDB) version 20201008.} following the standard guidelines from \cite{Brown09}. By visual inspection, we did not discern any contamination coming from the host galaxy, therefore we deemed unnecessary to perform any template subtraction. We did not detect SN~2022joj in the \textit{UV-W1} and \textit{UV-M2} filters during the first epoch, suggesting strong obscuration of the bluer part of the SED at early phases, like shown by the classification spectrum. 

\begin{deluxetable}{ccccc}
\tablewidth{0pt}
\tablehead{
\colhead{Filter} & \colhead{$t_{\rm{max}} (\rm{JD})$} & \colhead{$M_{\rm{max}}$} & \colhead{$\Delta m_{15}$}  
}
\startdata 
B & 2459722.3 & $-19.35 ^{+1.03}_{-0.65}$ & 1.4 \\                
g & 2459722.47 & $-19.52^{+1.04}_{-0.68}$ & 1.3 \\
V & 2459723.8 & $-19.32^{+1.03}_{-0.71}$ & 0.85 \\
r & 2459725.63 & $-19.27^{+1.04}_{-0.7}$& 0.75 \\
i & 2459721.94 & $-18.67^{+1.04}_{-0.7}$ & 0.77 \\
\enddata
\caption{\label{tab:table-name} Time of maximum light, absolute magnitude, and decline rate ($\Delta m_{15}$, the drop in magnitudes between peak and 15 days after peak) for SN~2022joj in each filter of its lightcurve. }
\label{table:1}
\end{deluxetable}

\subsection{Spectra}
\begin{figure}
\subfloat{%
  \includegraphics[clip,width=\columnwidth]{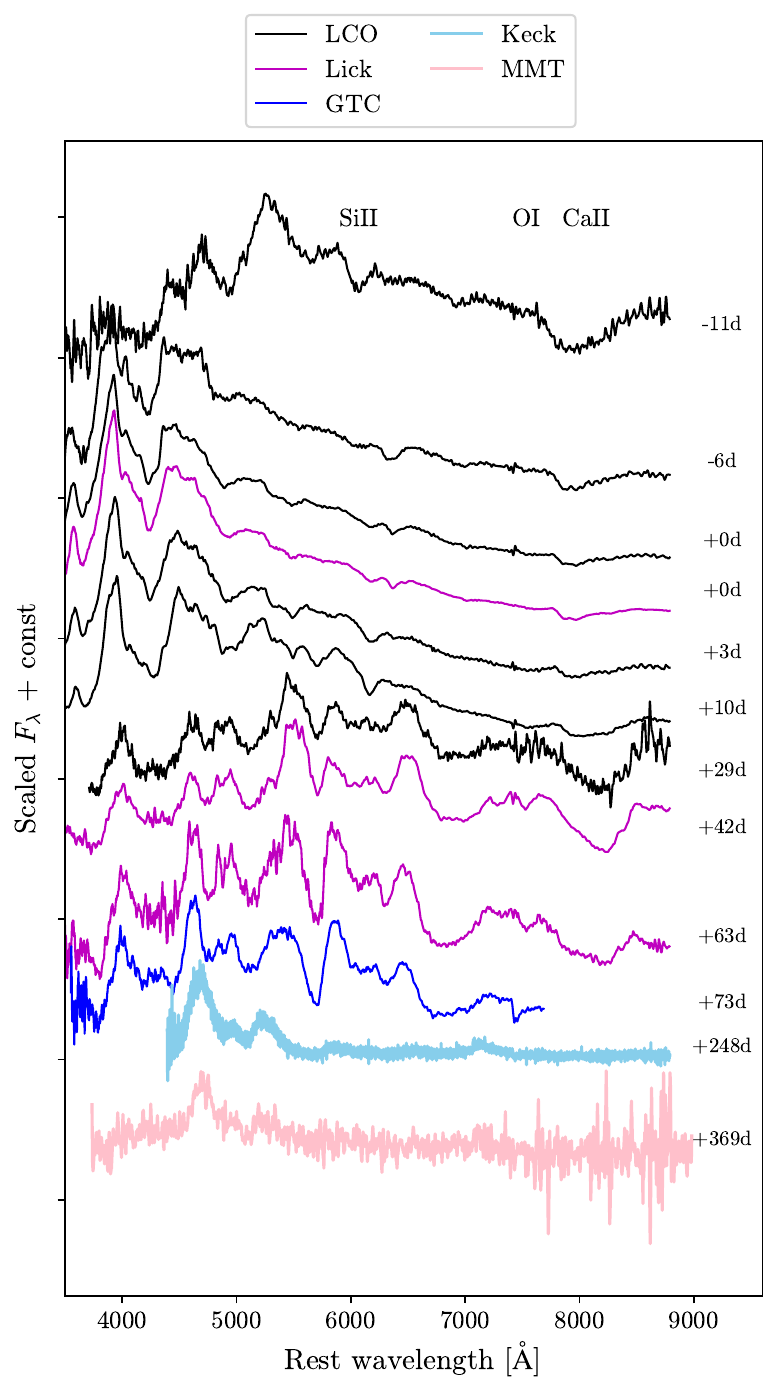}%
}
\vspace{-1.07\baselineskip}

\subfloat{%
  \includegraphics[clip,width=1.05\columnwidth]{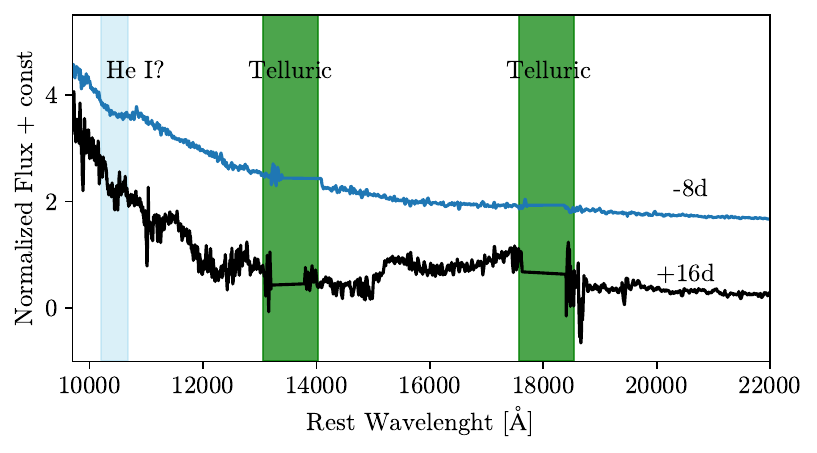}%
}
\caption{Milky Way extinction corrected spectra using $E(B-V)_{\rm{MW}}=0.0313$ mag at various phases with respect to maximum light in the B band. Note, these spectra are not corrected for host extinction (see Section \ref{sec:3.1} for details on host extinction).
\label{fig:22joj_spec}}
\end{figure}

The full spectroscopic data set is shown in Figure \ref{fig:22joj_spec}. A follow-up sequence was initiated from the Global Supernova Project using the FLOYDS spectrographs mounted on the 2m Faulkes Telescope North in  Haleakalā, Hawai'i, as well as the Faulkes Telescope South in Siding Spring, Australia. The data were reduced as detailed in \cite{Valenti14}. The redshift was determined by comparing the characteristics of the observed Type Ia supernova using Supernova Identification (SNID) fitting algorithm at maximum light \citep{Blondin07}. The best match at maximum light was found with SN 2000cx, resulting in a redshift of $0.024$ $\pm$ $0.009$. The redshift error is related to the width of the correlation peak and the \textit{rlap} parameter, which is used to quantify the reliability of a given correlation between
the input and a template spectrum \citep{Blondin07}. To get a sense of the dispersion of matches obtained from SNID, we examined the results of 76 potential candidates. Among them, 46 were identified as \textquote{good} matches, and those were selected to conduct a statistical analysis. We found the mean redshift to be 0.0225 with a standard deviation of 0.0057. This range is consistent with the best match at maximum light and therefore chosen. All spectra were corrected for Milky Way reddening, $E(B-V) =0.0313$ mag, as discussed in Section \ref{sec:discover}. We assume the host extinction to be negligible (see Section \ref{sec:remote_location}).

Multiple optical spectra were obtained from the Kast spectrograph at Lick Observatory, on 2022 May 22, June 29 and July 24. One additional spectrum was obtained from the Gran Telescopio Canarias (GTC), at the Observatorio del Roque de Los Muchachos in La Palma on 2022 August 3, taken with the Optical System for Imaging and low-Intermediate-Resolution Integrated Spectroscopy (OSIRIS). The reduction process was performed using version 1.11.0 of \textit{PypeIt} \citep{pypeit:joss_pub}\footnote{\url{https://github.com/pypeit/PypeIt}}. We secured a nebular spectrum from Keck+DEIMOS on 2023 January 17. \textit{PypeIt} was used for the reduction of this latter. A second nebular spectrum was obtained with MMT+Binospec \citep{Binospec} on 2023 May 26. This was reduced automatically by the Binospec IDL pipeline \citep{Kansky19}.
Near infrared data (NIR) from Keck taken on 2022 May 14, using the Near-Infrared Echellette Spectrometer \citep[NIRES;][]{Wilson2004} on the Keck~II telescope. The spectrum was reduced using the Spextool software package \citep{Cushing2004}. We also acquired data from Keck Infrared Transient Survey (KITS) collaboration on 2022 June 7 also from NIRES. 
These findings are visually represented in Figure \ref{fig:22joj_spec}, where the bottom plot showcases the NIRES data, and the top figure displays the optical and nebular data in ascending order.


\section{Analysis}\label{sec:Analysis}
\subsection{Light Curves and Colors} \label{sec:3.1}

\begin{figure*}
\centering
\includegraphics[width=0.95\textwidth]{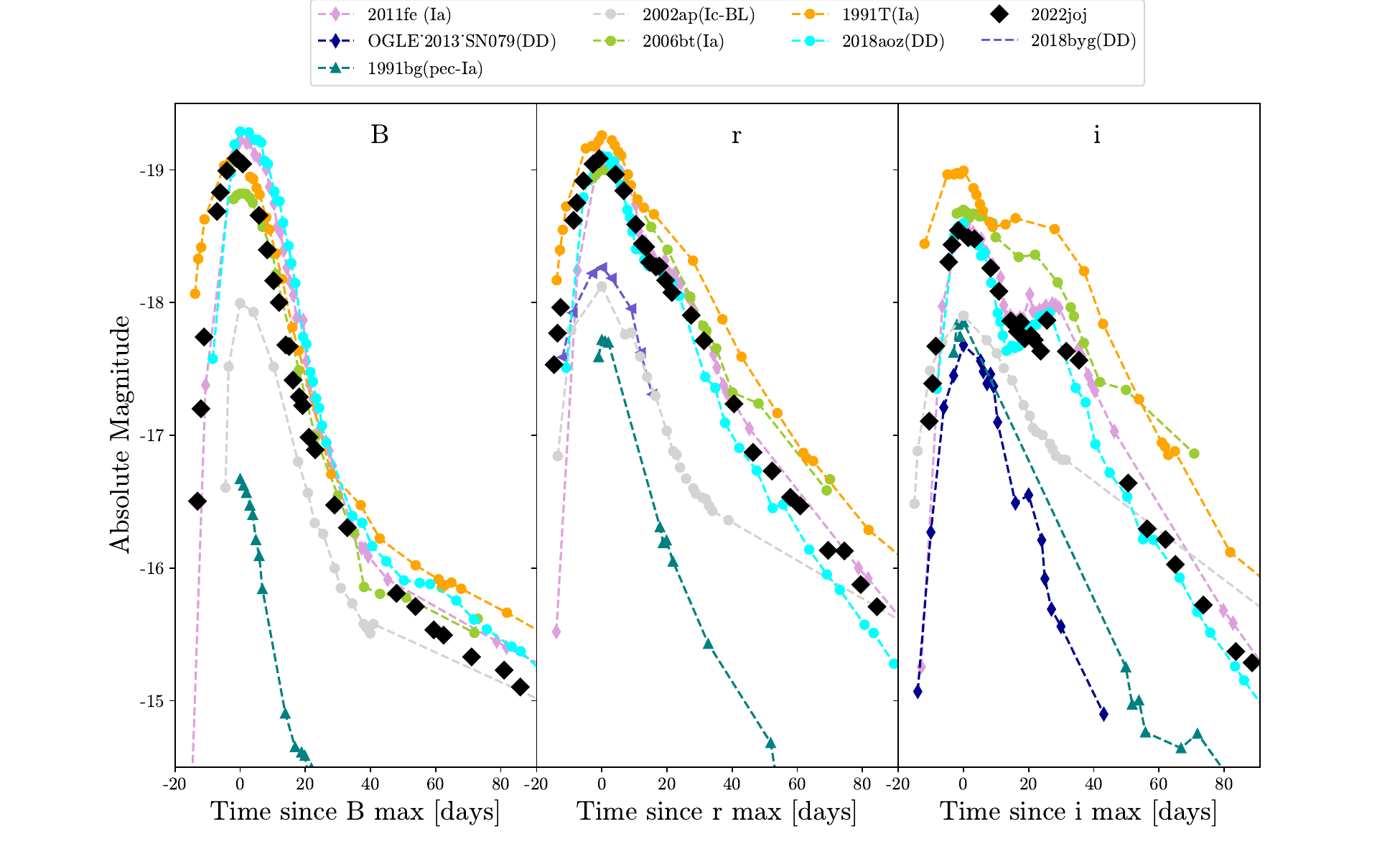}
\caption{Light curve comparison of SN~2022joj to other SNe Ia. Note that SN~2022joj declines faster than the typical SN Ia 2011fe. All of the light curves have been corrected for extinction. Note that SNe with \textquote{DD} next to their name in the legend correspond to double detonation candidates.
\label{fig:lc_com}}
\end{figure*}

In order to better understand SN~2022joj's evolution over time, we present a comprehensive comparative light curve analysis (Figure \ref{fig:lc_com}). The comparison encompasses the standard Type Ia, subluminous Type Ia examples like SN 1991bg \citep{filippenko1992} and SN 2002es \citep{Ganeshalingam2012}, as well as potential double detonation candidates including SN 2016hnk \citep{Galbany2019}, OGLE 2013 SN 079 \citep{Inserra_2015}, SN 2016dsg \citep{Dong2022}, SN 2018aoz \citep{Ni2022}, SN 2018byg \citep{De_2019}, and SN 2019eix \citep{PadillaGonzalez2022}. Additionally, it showcases an atypical broad-line Type Ia, SN 2002bo \citep{OBrien2021} and draws a comparison to an energetic Broad-Line Type Ic, SN 2002ap \citep{Mazzali2002}. Note that all SNe are Milky Way and host extinction corrected. 

\begin{figure}
\includegraphics[width=0.52\textwidth]{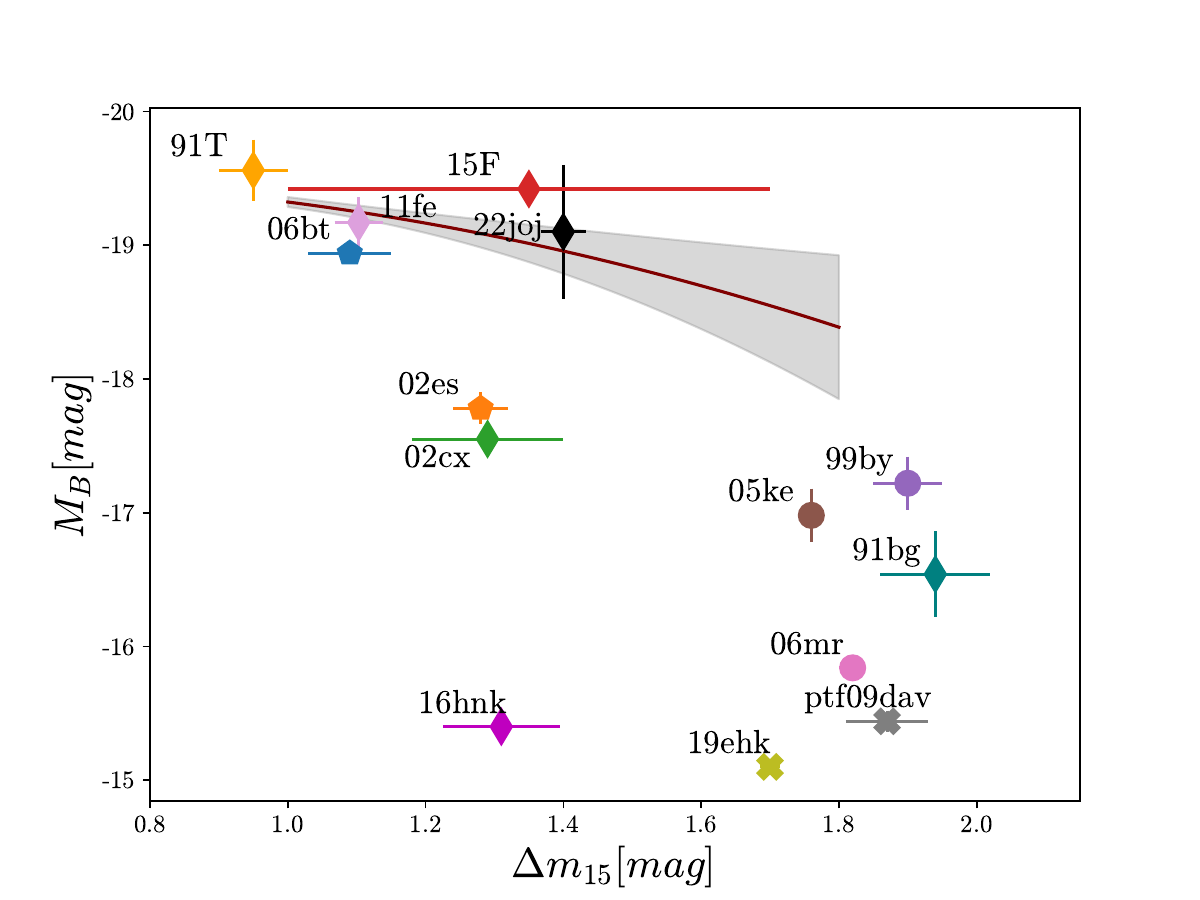}
\caption{Phillips relation is shown in the brown line along with various peculiar SNe. Note how SN~2022joj (in black) agrees in relation to its lightcurve width. The symbols used to represent these different types are: normal SNe Ia (diamonds), 91bg-like SNe Ia (circles), SNe Iax (stars), 02es-like SNe Ia (pentagons), Ca-rich objects (crosses), SN 2016hnk (down triangle), and overluminous SNe Ia 1991T (right triangle).
\label{fig:philips_rel}}
\end{figure}

Figure \ref{fig:lc_com} illustrates that SN~2022joj exhibits a similar peak absolute magnitude compared to both the standard SN Type Ia and the double detonation with a thin He-shell candidate, SN 2018aoz. Note that the peak luminosity has high uncertainties of $\pm$ 1 mag, due to the uncertainty of the redshift (and potential unaccounted reddening from host). Additionally, from Figure \ref{fig:philips_rel}, it is evident that the light curve's width is typical of SNe Ia behavior. SN~2022joj follows the Phillips relation that links the brightness and decline rate of Type Ia supernovae \citep{Phillips1993}. Moreover, we observe that the decline rate is swifter than that of SN 2018aoz, indicating a possible less massive progenitor. These diverse observations could also be angle dependent, as discussed in Section \ref{sec:Modeling}.

The lightcurves of SN~2022joj, as shown in Figure \ref{fig:lc_com}, reveal a subtle secondary maximum in the i band, a characteristic often observed in sub-luminous SNe. This behavior is attributed to their cooler supernova photospheres \citep{Foley_2009, Kasen2006}, which offer valuable information about the temperature of supernovae. In Figure \ref{fig:color_evol}, we compare the color evolution of SN~2022joj with that of typical and atypical SNe Ia, as well as double detonation candidates and models. 


\subsection{Spectral Analysis} \label{sec:3.2}
\begin{figure*}
\centering
\includegraphics[width=0.9\textwidth]{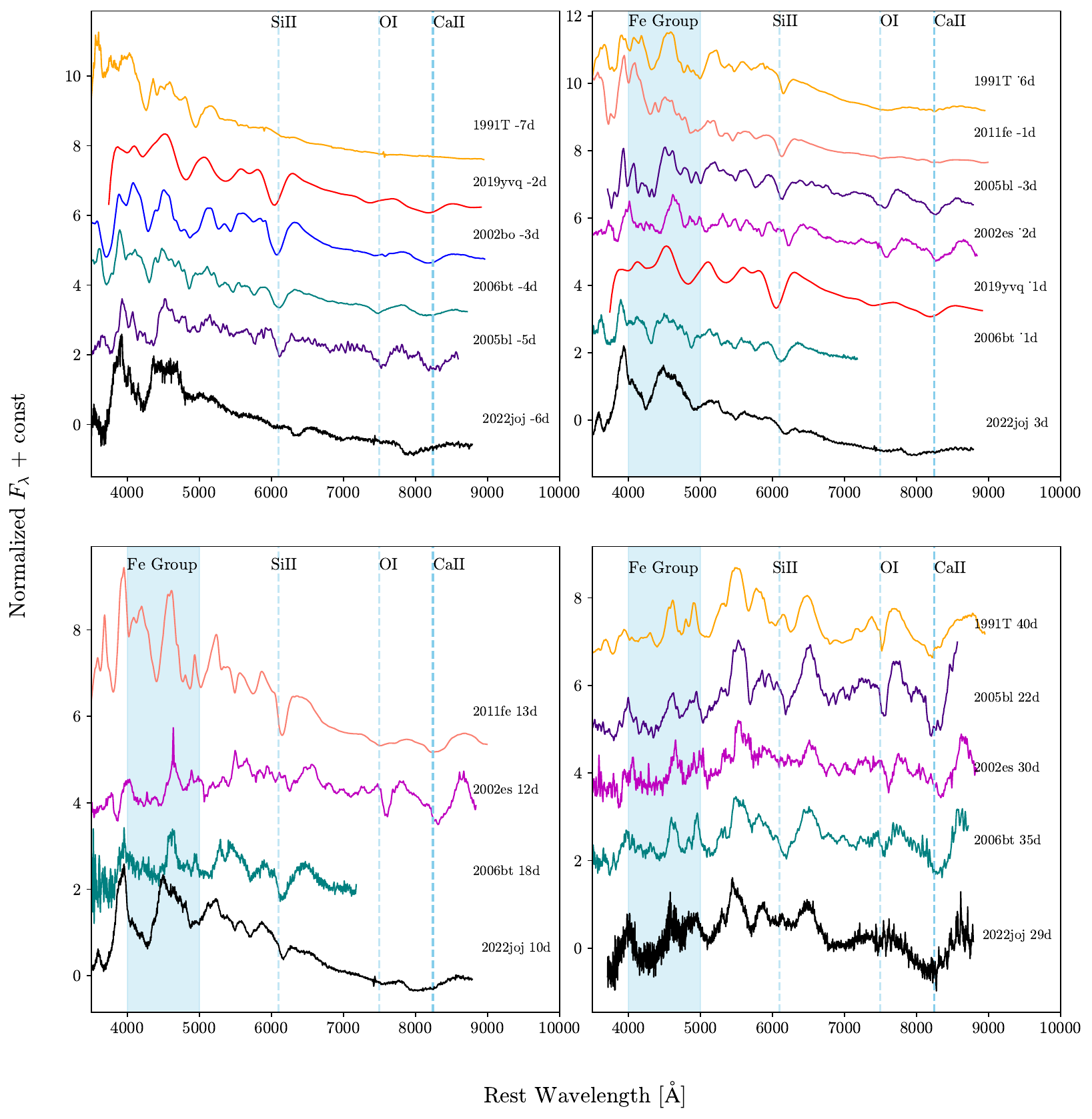}
\caption{Spectral evolution of SN~2022joj compared with other SNe Ia. The light blue dashed lines show the absorption features created by the respective line labeled at the top of the figure. Note how SN~2022joj resembles subluminous SN spectra despite being brighter. The top left panel shows SN~2022joj at early phases. The top right panel shows the spectra of SN~2022joj and other SNe Ia near maximum light. The bottom panels show SN~2022joj after maximum light. Note how SN~2022joj has weak \ion{Si}{2} and \ion{O}{1} feature throughout its evolution and develops strong \ion{Ca}{3} triplet.
\label{fig:spec_comp}}
\end{figure*}

The spectra of SN~2022joj were analyzed in comparison to several distinctive objects, primarily due to its remarkable features apparent in its initial phase. The most prominent of these features was the suppression of the blue wavelengths ($\lambda < 5000$\AA), setting it apart from other objects as shown in Figure \ref{fig:spec_comp}. As SN~2022joj evolved, it began to resemble a standard Type Ia. However some features commonly observed in a typical SN Ia, such as the \ion{Si}{2} 6355 feature, were not prominent in SN~2022joj. Additionally, we also observe an absorption minimum of the \ion{Ca}{2} triplet at a much higher velocity than the other objects throughout its evolution as shown in Figure \ref{fig:spec_comp}.

SN~2022joj exhibits characteristics \textquote{similar} to those of sub-luminous SNe Ia approximately 5 days before reaching maximum brightness (minus the deep \ion{Si}{2} 6355 feature and the lack of \ion{O}{1} 7777), as evidenced by the presence of \ion{Ti}{2} in the $4000< \lambda < 4500$ $\rm{\AA}$ range. This could be evidence of a double detonation of a He-shell, since Ti is predicted to be synthesized in the high-velocity outer layers of the ejecta \citep{Fink10, Jiang2017} along with the high velocities of \ion{Ca}{2} triplet, resulting from the detonation of the He shell \citep{Fink10,Kromer2010,Moore_2013}. Additionally, SN~2022joj appears to have a weak \ion{Si}{2} 6355 and unlike the double detonation models, it appears to have a strong \ion{C}{2} 6580 as shown in Figure \ref{fig:spectra_ddmodels}. Similarly, in the early epochs, SN 2005bl established a presence of \ion{C}{2}, suggesting an overall low burning efficiency with a significant amount of leftover unburned material \citep{tauberger2008}. However, \cite{Blondin2018} attributed this feature to absorption by the \ion{Mg}{2} 6347 doublet absorption in their sub-Chandrasekhar model.

At around maximum brightness, SN~2022joj continues to resemble sub-luminous SNe Ia spectroscopically. We also notice that the \ion{Si}{2} 6355 feature dominates over the \ion{C}{2} 6580 feature previously seen. Moreover, SN~2022joj shares a striking resemblance with peculiar SN 2006bt, as both share similarities with those of low-luminosity SNe Ia. Roughly two weeks to a month after maximum light, the \ion{Si}{2} 6355 feature remains significantly weaker in SN~2022joj than in SN 2011fe. At the one-month mark after maximum light, SN~2022joj bears a strong resemblance to the sub-luminous SN 2005bl, which displays a broad absorption trough between wavelengths of $4000$ to $4500$\AA, resulting from a blend of Fe-group elements dominated by Ti II \citep{filippenko1992, mazzali1997}.

\section{Modeling} \label{sec:Modeling}

We compared SN~2022joj with double detonation models from the existing literature due to its initial spectral peculiarities and color characteristics. We conducted photometric and spectroscopic analyses, comparing SN~2022joj with a variety of models, including those proposed by \cite{Kromer2010}, \cite{Polin2019}, \cite{kenshen2021}, and \cite{Ni2022}. In particular, \citet{Kromer2010} investigated the observable properties of double detonation models, which were simulated using 2D radiative transport code \texttt{ARTIS}. These models were initialized with estimated values of temperature, central density of the CO core, and temperature and density at the base of the He layer. An initial He detonation was ignited at a single point at the base of the He shell, eventually creating a shock wave that propagated and converged into the core. The He shell masses in these models, which had previously been considered by \cite{Fink10}, ranged from 0.0035 to 0.0126 $\rm M_{\odot}$ and WD masses spanning $0.8 - 1.38$  $\rm M_{\odot}$.

\citet{Polin2019} investigated the explosions of white dwarfs with He shell masses of 0.01, 0.05, and 0.08 $\rm M_{\odot}$ and WD masses ranging from 0.6 to 1.2 $\rm M_{\odot}$. Thicker shell models were found to exhibit early time flux excess, redder colors, and higher line blanketing in the UV through the blue regime of the spectrum. To create these 1D models, the authors employed the Eulerian hydrodynamics code \texttt{Castro}. Once the SN ejecta reached homologous expansion, synthetic spectra and light curves were generated using the multi-dimensional time-dependent radiation transport code \texttt{SEDONA}.

The models presented in \cite{Boos2021} and \cite{kenshen2021} utilized two-dimensional sub-Chandrasekhar-mass double detonation models as initial parameters, which were then input into the reactive hydrodynamics code \texttt{FLASH}. To initiate the explosion in each model, a hotspot was placed along the helium shell symmetry axis. This resulted in the helium detonating around the surface to the south pole, generating a shock wave that propagated into the core and ultimately triggered a carbon-core detonation. The explosion parameters in these models varied, with core masses ranging from 0.82 to 1.09 $\rm M_{\odot}$ and shell masses between 0.011 and 0.1 $\rm M_{\odot}$. The shell masses were not comprised solely of $^{4}\rm{He}$, but rather were mixed with $^{12}\rm{C}$, $^{4}\rm{N}$, and $^{16}\rm{O}$.

Similar to \cite{Polin2019}, \cite{Ni2022} He-shell double detonation 1D models, hydrodynamics, and nucleosynthesis simulations were conducted using \texttt{Castro} and the radiative transfer calculations were conducted using \texttt{Sedona}. The parameter space covered for core masses ranged from 1.0 to 1.1 $\rm M_{\odot}$ and the shell masses ranged from 0.01 to 0.012 $\rm M_{\odot}$.

\begin{figure*}
\centering
\includegraphics[width=0.5\textwidth]{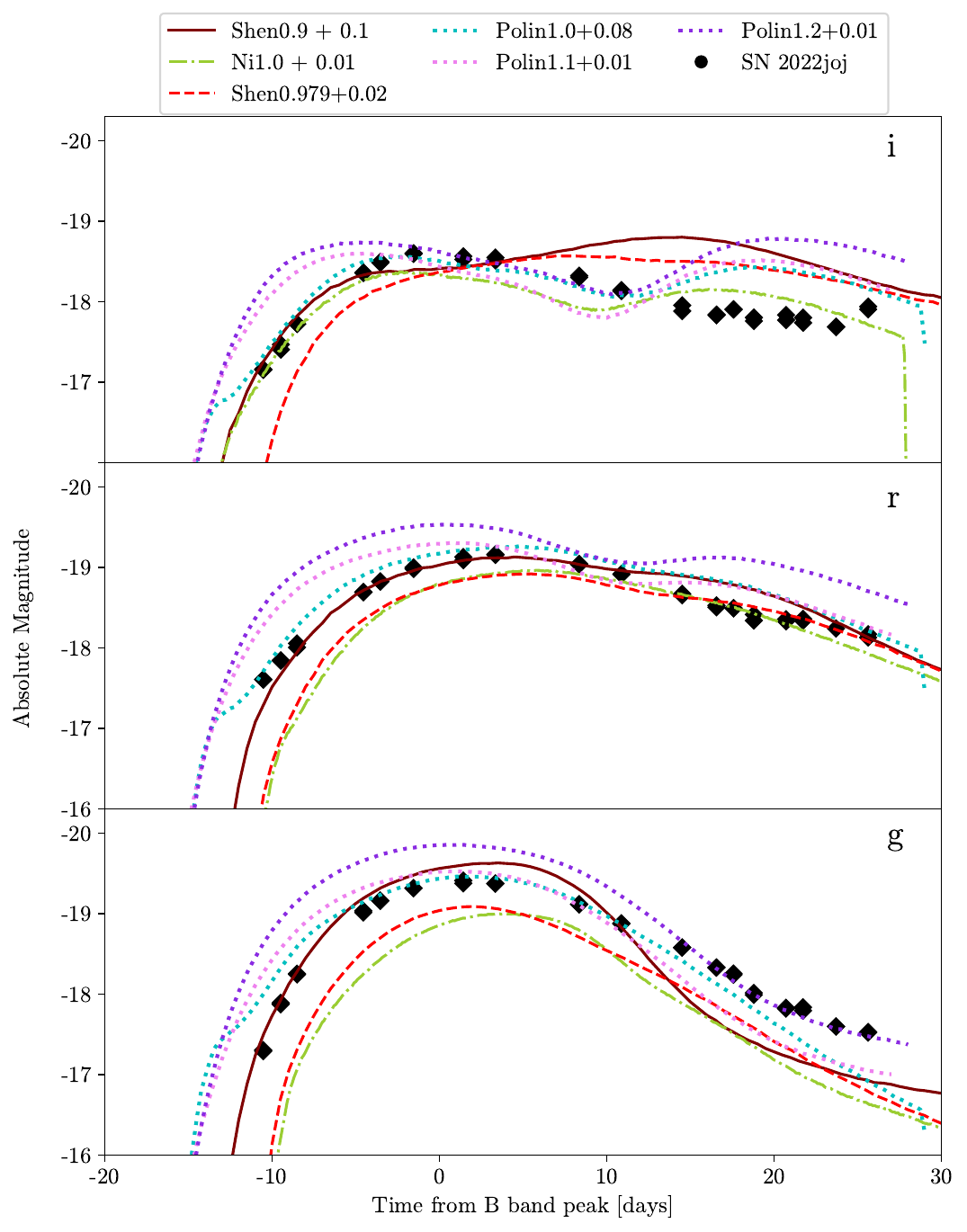}
\caption{Lightcurve comparison between SN~2022joj and double detonation models with a thin and thick He-shell. The phase is measured from the B band maximum. 
It is noteworthy that the majority of models, especially Shen 0.9+0.1 and Polin 1.0+0.08, effectively capture the light curves of SN~2022joj.
\label{fig:lc_ddmodels}}
\end{figure*}

\begin{figure*}
\centering
\includegraphics[width=0.95\textwidth]{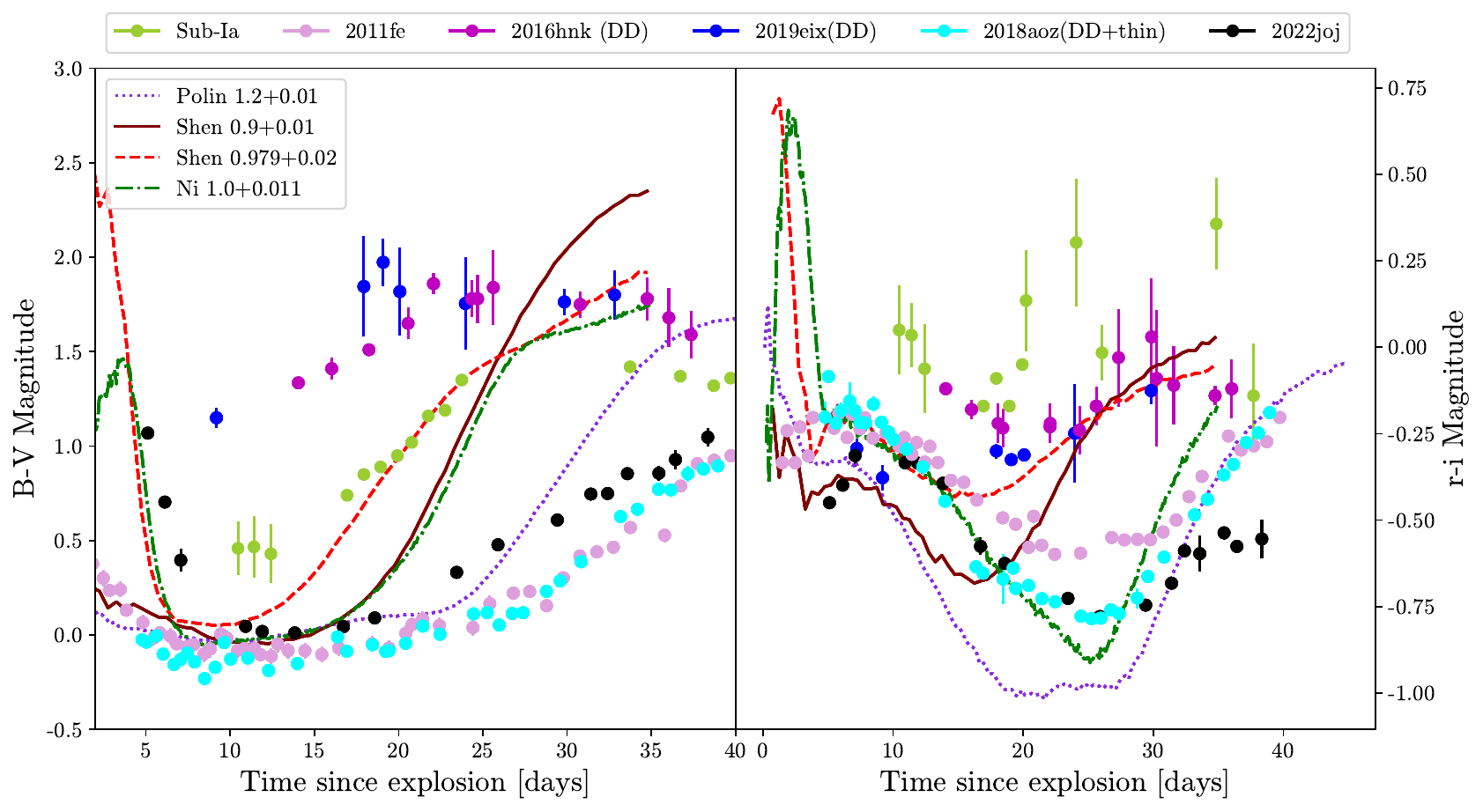}
\caption{The color evolution of SN~2022joj plotted against other SNe in terms of explosion time, where our explosion time was calculated using a polynomial fit. We plotted double detonation models (dashed lines), where the green correspond to 1.0+0.011 $\rm{M_{\odot}}$ from \cite{Ni2022}, the purple are 1.2+0.01 $\rm{M_{\odot}}$ from \cite{Polin2019}, and the maroon and red are \cite{kenshen2021} 0.9+0.1 and 0.979+0.02 $\rm{M_{\odot}}$. Additionally, the \textquote{Sub-Ia}category, denoting sub-luminous Type Ia supernovae, is illustrated with data points representing SN 1991bg and SN 2005bl. Note how SN~2022joj is much redder at early phases in comparison to all of the SNe, especially SN 2011fe.
\label{fig:color_evol}}
\end{figure*}

\subsection{Light curve model analysis} \label{sec:lc modeling}
In Figure \ref{fig:lc_ddmodels}, we plot SN~2022joj light curves in the $i$, $r$, and $g$ bands against various double detonation models with thin and thick He shells. We note that the SN~2022joj light curve width and magnitude aligns with several of them, including Shen 0.9+0.1 and Polin 1.0+0.08. However, it is important to note that SN 2022joj has a large uncertainty in its absolute magnitude due to its distance uncertainty. Moreover, models assume local thermal equilibrium (LTE) after maximum leading to additional discrepancies, specifically a too-fast decline in B band magnitude compared to more realistic non-LTE models \citep{kenshen2021}. Additionally, it is interesting to see WD masses in that mass range, since C/O white dwarfs at formation are thought to be limited to a maximum mass of  $1-1.1$ $\rm{M_{\odot}}$ \citep{Dominguez1999,Girardi2002, Catal2008}. The nature of such a massive progenitor is an open question. 

One possibility is that it grew by stable surface burning \citep[H or He to C and/or O;][]{Wolf_2013}, though how the progenitor could go from stable burning to a helium shell detonation is unclear. A common channel found in binary population synthesis involves stable helium mass transfer from a helium star on a thermal timescale, growing the C/O white dwarf by a couple of tenths of a solar mass. The helium star then evolves into another C/O white dwarf, and their merger can lead to a double detonation event \citep{Ruiter2013}. Another channel could be a hypothesized hybrid C-O-Ne white dwarf \citep{2014MNRAS.440.1274C}, sometimes invoked as the progenitors of SNe Iax \citep{2014ApJ...789L..45M}. However, the \citet{Polin2019} and  \citet{kenshen2021} models did not use C-O-Ne progenitors.

SN~2022joj stands out due to its extreme redness in the early stages. However, as it evolves, it appears to become more similar to a typical Ia such as SN 2011fe. We explored the possibility of a double detonation involving a thin He-shell as shown by \cite{Shen18}, \cite{Townsley2019}, \cite{Boos2021}, \cite{kenshen2021}, \cite{Ni2022}, and \cite{collins2022} that such events can reproduce typical SN Ia features. In Figure \ref{fig:color_evol}, we plot color evolution of SN 2022joj along with double detonation models. Our analysis indicate that SN~2022joj is much redder at early epochs than SN 2011fe. Notably, the Ni 1.0 + 0.01 (referring to a 1.0 $M_{\odot}$ WD mass with a helium shell of 0.01 $M_{\odot}$) and Shen 0.0979+0.02 models, accurately predict the colors of SN~2022joj, despite showing a redder trend slightly earlier than depicted in the observations. However, this can be attributed to the uncertain explosion time of SN~2022joj.

\subsection{Spectra model analysis} \label{sec: spectra modeling}
\begin{figure*}
\centering
\includegraphics[width=0.85\textwidth]{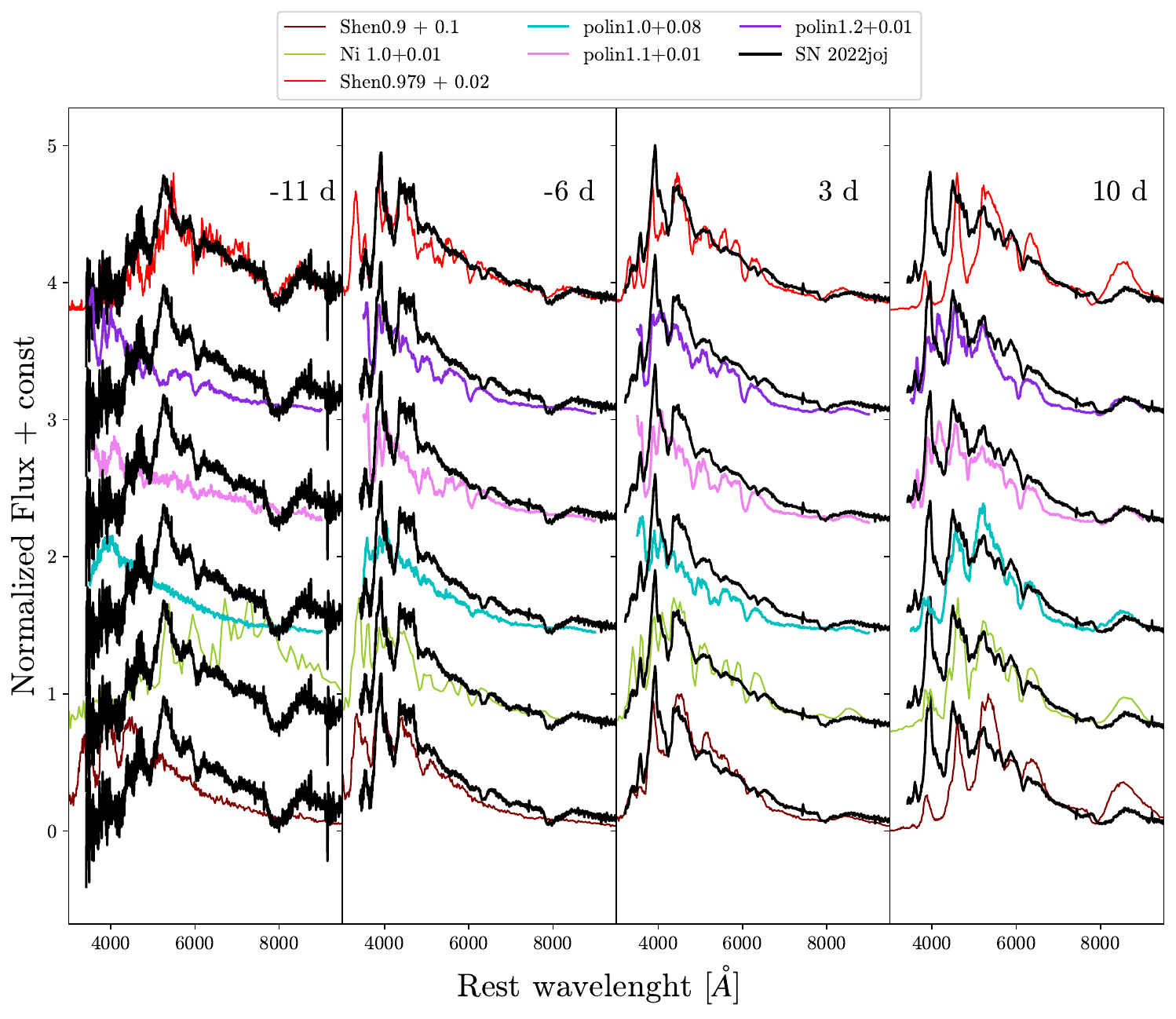}
\caption{Spectroscopic comparison between SN~2022joj and double detonation models. The left panel shows the spectra at 11 days before maximum light. SN~2022joj shows an extreme reddening and is best matched with Shen 0.979+0.02. The second and third panels correspond to 6 days before maximum light and near maximum light. These two panels are best matched with Shen 0.979+0.02, Shen 0.9+0.1, and Polin 1.2+0.01. Lastly, the right panel is 10 days after maximum light that seems to be best matched with Polin 1.2+0.01. 
\label{fig:spectra_ddmodels}}
\end{figure*}

Figure \ref{fig:spectra_ddmodels} illustrates the spectral profiles of SN~2022joj both before and after reaching maximum brightness, juxtaposed with thin and thick He-shell double detonation models mentioned in the literature. Shen's 0.9+0.1 and 0.979+0.02 models exhibit striking similarities in spectra with SN~2022joj, including \ion{Ti}{2} absorption, a shallow \ion{Si}{2} feature, and a lack of \ion{O}{1} absorption features ranging from six days before maximum light to maximum light. However, discrepancies after maximum light can be attributed to the models assuming LTE, resulting in lower temperatures and more singly ionized Fe-group material \citep{shen2021nonlte}. According to \cite{kenshen2021}, the \ion{Ti}{2} absorption is due to the lower temperatures caused by both the more radially extended distribution of Ti and the lower flux along these lines of sight. Moreover, Shen 0.979+0.02 effectively reproduces the early spectrum at $-11$ days before maximum light.

Nevertheless, despite Shen's models considering He mixing in the shell with C, the \ion{C}{2} 6580 feature at $-6$ days before maximum light is not reproduced by their models. This discrepancy might suggest that SN~2022joj contains a significantly higher amount of C in the shell compared to what the models have predicted.  By replicating the primary characteristics that distinguish SN~2022joj from typical SN Ia, this model is successful in demonstrating its uniqueness. The differences in the observables of \cite{kenshen2021} to the other models can be attributed to the fact that it was a 2D simulator which can account for various viewing angles. We found that the best viewing angle ($\mu$, defied as $\cos (\theta)$) that matches SN 2022joj is at $\mu = -0.93$ for the 0.9+0.1 model and $\mu = +0.93$ for the 0.979+0.02 model. Notably, $\mu = -0.93$ is observed in the southern hemisphere where the carbon-core detonation is ignited, while $\mu = +0.93$ is observed from the northern hemisphere, where the He is ignited. It is important to note that these chosen matches have drastically different shell thicknesses (at opposite lines of sight), which demonstrates the importance of the multidimensional aspect of the double detonation model. Overall, we find that Shen 0.979+0.02 model best matches SN~2022joj spectral peculiarities.

\subsection{Nickel Distribution Analysis}
Due to the peculiar colors and brightness illustrated in the light curves, we investigate the effects on various nickel distributions. \cite{Magee2018}, found that for a given density, models with $\rm{^{56}Ni}$ extending through the ejecta were brighter and bluer at earlier times than models in which $\rm{^{56}Ni}$ was concentrated. Their models were calculated using TURTLS, a radiative transfer code where the density profile is altered. For each density profile, a series of $\rm{^{56}Ni}$ distributions was used, which decrease towards larger radii as shown in equation \ref{eqn:nickel_eqn}. The scale parameter \textit{s} controls how quickly the ejecta transitions from $\rm{^{56}Ni}$-rich to $\rm{^{56}Ni}$-poor, where a larger \textit{s} represents a sharper transition between the two regions. These scaling parameters ranged from 3 to 100 and the  $\rm{^{56}Ni}$ masses from 0.4 to 0.8 $\rm{M_{\odot}}$, covering the expected range for typical SNe Ia \citep{magee2020}.

\begin{equation}
\label{eqn:nickel_eqn}
{^{56}{\rm Ni}(m)} = \frac{1}{\exp\left(s\left[m-M_{\rm{Ni}}\right]/M_{\odot}\right)+1}
\end{equation}

\begin{figure}
\centering
\includegraphics[width=0.5\textwidth]{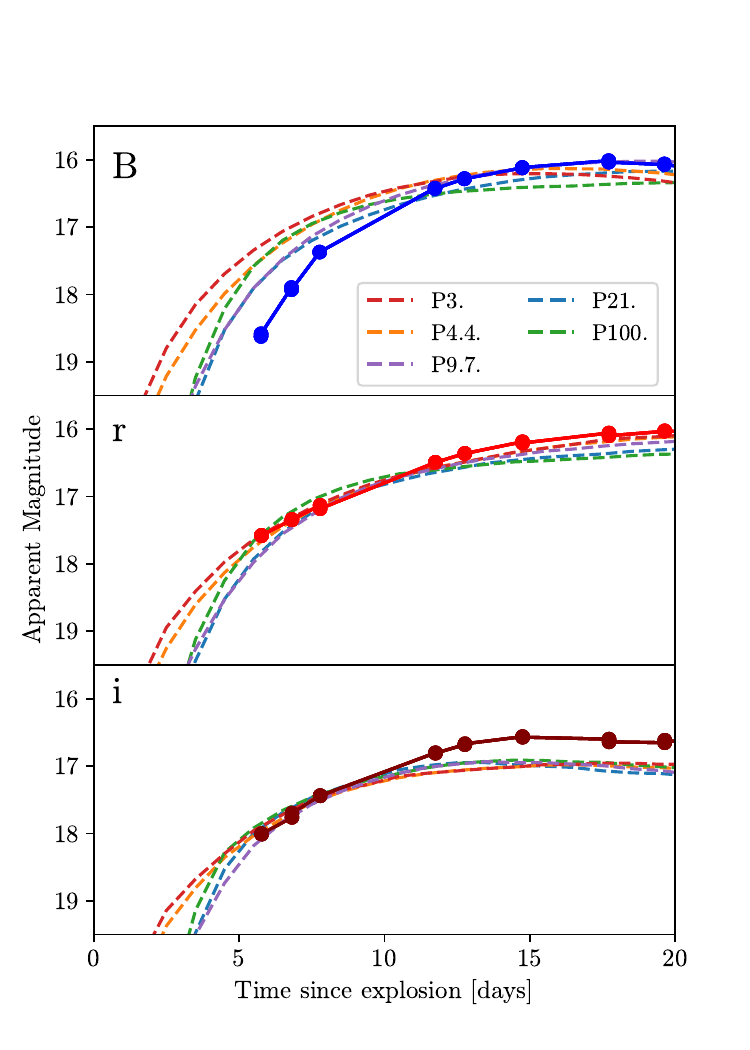}
\caption{SN~2022joj in the B, r, and i band amongst various Ni-distributions from \cite{magee2020}. The models assume a Ni mass of 0.8 $\rm{M_{\odot}}$ and a kinetic energy of $1.4\times10^{51}$ erg. The different values of P correspond to the scaling parameter, where the larger values indicate sharper transitions from Ni-rich to Ni-poor. 
\label{fig:ni_dist}}
\end{figure}

In Figure \ref{fig:ni_dist}, we plotted the best fit models for each scaling parameter with the minimum $\chi^{2}$. From the figure we notice that the lower the scaling parameter the broader and brighter the early light curve is, due to the $\rm{^{56}Ni}$ being farther out than those with a larger scaling parameter (where most of the $\rm{^{56}Ni}$ is close to the core). SN~2022joj has a best fit when the scaling parameter is 9.7. Although the light curves are a decent fit, the $B$ band is highly overestimated, indicating that shallower $\rm{^{56}Ni}$ distributions predict bluer colors at the early phases, which is the opposite for SN~2022joj. Therefore, a shallower $\rm{^{56}Ni}$ cannot explain the early colors seen in SN~2022joj.

\subsection{NIR analysis}\label{sec: NIR analysis}

Recent simulations have indicated the presence of unburned helium in both single and double detonations in the outer ejecta \citep{Fink10,Shen10, Kromer2010, sim12, Polin2019, collins2022}. However, the studies could not fully address the question of whether helium spectral features should form in the models owing to approximations used in the atomic physics in both studies. In \cite{dessart&Hillier14}, involving non-LTE simulations, it was found that the spectral lines from unburned helium, specifically the \ion{He}{1} $\lambda$ 10830 line, can be observed in the single detonation. The \ion{He}{1} $\lambda$ 10830 line exhibits a P-cygni profile and can be visible up to 5 days after the explosion. However, the light curves projected by their model are over 2 magnitudes dimmer than SN~2022joj. This is because the models were formulated for single detonation events, which results in fainter observables.

\begin{figure}
\includegraphics[width=0.5\textwidth]{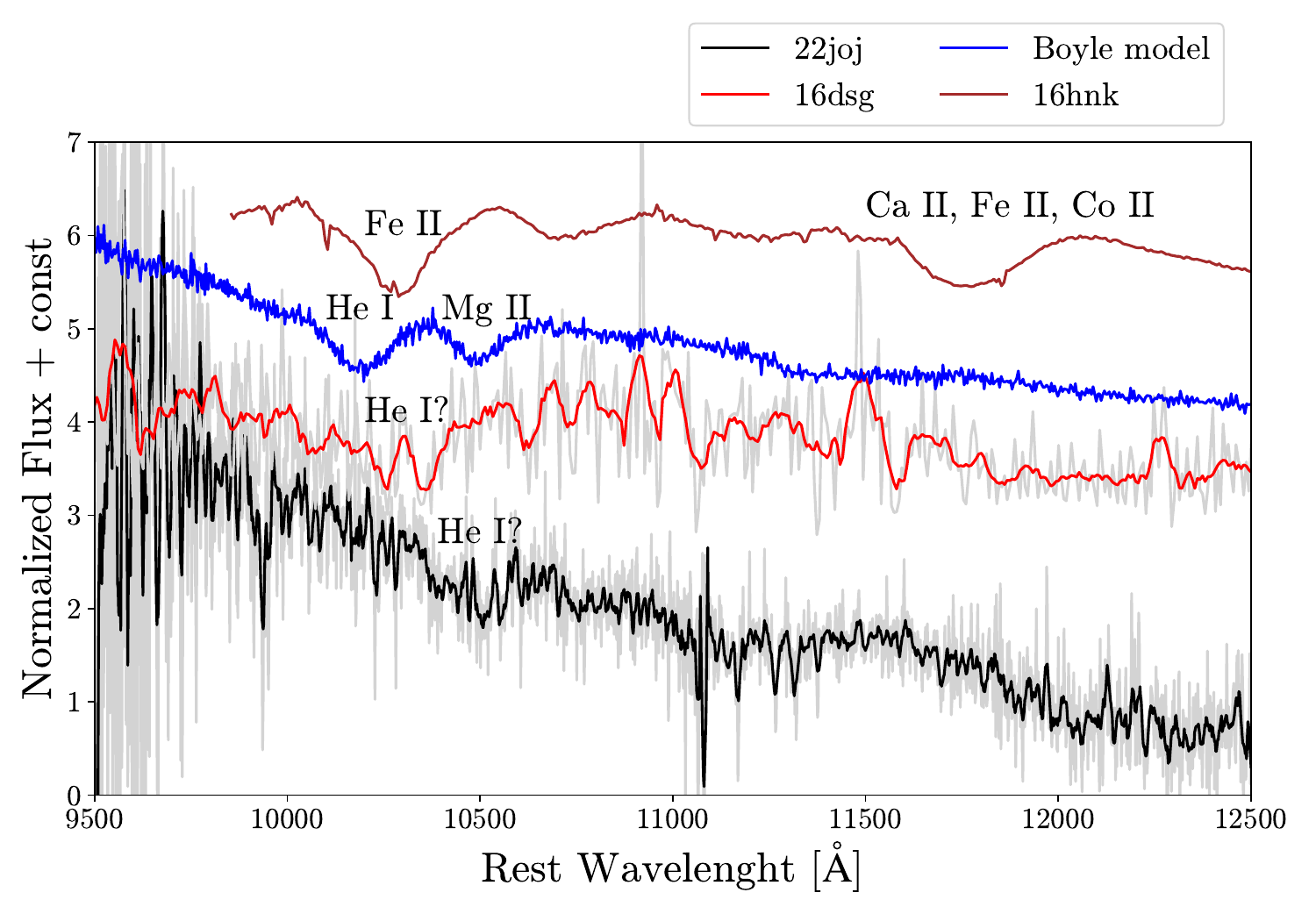}
\caption{NIR spectra of SN~2022joj plotted against other double detonation candidates and a NIR model from \cite{Boyle2017}. Note that the model assumes a He-shell mass of 0.05 $\rm{M_{\odot}}$ and that the \ion{He}{1} absorption feature is more apparent than in the observations. Moreover, SN~2022joj and SN 2016dsg are both plotted at +16 days after reaching maximum brightness, while Boyle's model and SN 2016hnk are depicted at +7 days following their respective peaks.
\label{fig:NIR_fig}}
\end{figure}

Nonetheless, the \ion{He}{1} $\lambda$ 10830 line has been explored by \cite{Boyle2017} who found that this line could be observed around maximum light and afterwards. In Figure \ref{fig:NIR_fig}, we show the NIR spectra of SN~2022joj, along with SN 2016dsg \citep{Dong2022}, SN 2016hnk \citep{Galbany2019}, and the high-mass model of Boyle at 26 days after explosion (+7d after maximum light). The purpose of this is to illustrate whether there exists unburned He for these double detonation candidates (SN~2022joj, along with SN 2016dsg, SN 2016hnk). In SN 2016dsg, it is suspected that the He line lies somewhere between 9700-10500 $\rm \AA$ \citep{Dong2022}. On the other hand, for SN 2016hnk, the absorption feature between 9700-10500 $\rm \AA$ is identified as \ion{Fe}{2}. Although SN~2022joj seems to be somewhat featureless, there does seem to be a dip around this range that could possibly correspond to the \ion{He}{1} feature. This velocity feature appears to be at a lower velocity compared to the model; however, this observation could be influenced by potential \ion{Mg}{2} blending. However, a study by \cite{Collins2023}, simulated full non-local thermodynamic equilibrium radiative
transfer models for a double detonation explosion model. At early epochs (5 days after explosion) they found that the \ion{He}{1} 10830 was blended with \ion{Mg}{2} 10927. However, this feature separates to form a secondary feature while becoming weaker over time. Therefore, we would expect that this feature found in SN 2022joj would have been separated according to \cite{Collins2023} models.

We also plotted the high-mass models from \cite{Boyle2017} based on a progenitor with a CO core mass of 1.025 $\rm M_{\odot}$, and a helium shell mass of 0.055 $\rm M_{\odot}$. The simulation resulted in 0.03 $\rm M_{\odot}$ of unburned helium remaining after the explosion and is plotted in Figure \ref{fig:NIR_fig}. It is not surprising that the \ion{He}{1} feature in the simulations from \cite{Boyle2017} is stronger than that in SN~2022joj. This difference can be attributed to their models, which assumed a He-shell of 0.05 $\rm M_{\odot}$, while the best simulations for SN~2022joj assumed a He-shell of 0.02 $\rm M_{\odot}$. 

\subsection{Nebular model analysis} \label{sec:nebular modeling}
The nebular spectrum is a powerful probe of the internal structure of SNe, including geometric asymmetries in the ejecta. \cite{Polin_2021} used the 1D hydrodynamic model \texttt{SedoNeb} which calculates the emissivities of each atomic transition by solving for the temperature, ionization state, and NLTE level populations. They used the \texttt{Sedona} models from \cite{Polin2019} as their input. The 1D models indicate that the prevailing patterns are primarily influenced by the occurrence and arrangement of $\rm{^{56}Ni}$ and $\rm{^{40}Ca}$. In particular, they found that low-mass double detonation models with only a small mass fraction of Ca produce nebular spectra that cool primarily through forbidden [Ca II] emission. The more massive progenitors produce spectra with strong Fe lines (4500–5600 \AA). However, even their brightest double detonation model overproduces [Ca II] $\lambda\lambda$ 7291, 7323 emission when compared to SNe Ia. 

\begin{figure}
\includegraphics[width=0.5\textwidth]{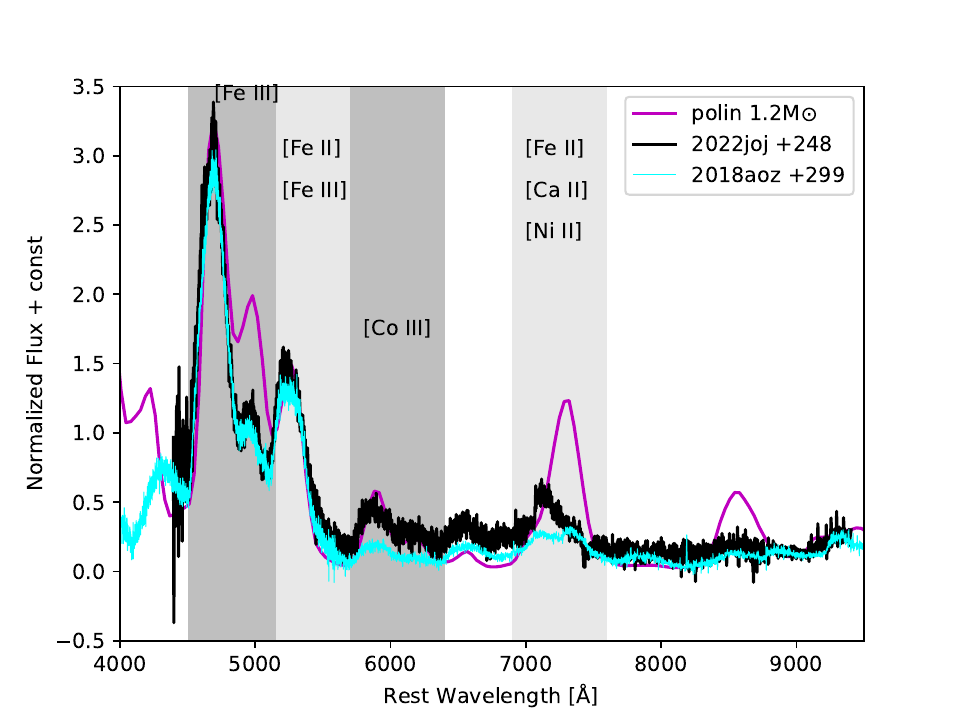}
\caption{Nebular spectrum of SN~2022joj, double detonation candidate SN 2018aoz, and the model spectra of Polin 1.1+0.01 $\rm{M_{\odot}}$. The model of the nebular spectrum predicts stronger [Ca II] emission than observed for SN~2022joj.
\label{fig:neb_ddmodels}}
\end{figure}

In Figure \ref{fig:neb_ddmodels} we show a comparison between SN~2022joj, SN 2018aoz (a double detonation with a thin He-shell candidate), and the modeled double detonation spectra at 150 days after maximum light. Due to the presence of a shallow [Ca II] feature in SN~2022joj, we performed a comparison with the shallowest [Ca II] from the nebular models presented in \cite{Polin_2021}. It was revealed that these models predict a significantly stronger [Ca II] feature than what was observed for both SN~2022joj and SN 2018aoz as shown in Figure \ref{fig:neb_ddmodels}. Notably, previous spectroscopic analyses had shown a decent match with Polin 1.2+0.01. Therefore, it was surprising to see the disagreement in the nebular phase. These uncertainties could be attributed to atomic data uncertainties and limitations in the 1D simulations; especially since double detonations observations can be highly viewing angle dependent \citep{kenshen2021}.

\section{Discussion} \label{sec:discussion}

\subsection{Implications for SN origins}
 SN~2022joj appears to obey the Phillips relationship that shows the correlation between brightness of the supernova and the rate of decline \citep{Phillips1993}. Additionally, SN~2022joj shows peculiarity in its colors and classification spectra. SNe Ia typically show a very blue continuum at these early phases. In SN~2022joj the colors exhibit an exceptional reddening 11 days prior to reaching maximum light, indicating the presence of a layer that absorbs blue light during these initial stages. 

When a WD undergoes a double detonation with a thin helium shell, the outer He layer burns to intermediate mass elements, which can produce strong UV line blanketing and cause a suppression in the blue side of the spectrum \citep{Fink10,Shen10, Kromer2010, sim12, Polin2019, Boos2021,collins2022}. We compared the light curves, spectra, and colors of these models to SN~2022joj. It is important to note that LTE models after maximum light are much redder than non-LTE due to low temperatures. Therefore, we observe a mismatch in colors and spectra after maximum light, but we include them for completeness \cite{shen2021nonlte}. The modeled spectra (especially at maximum light) agree nicely with our data. We found Shen 0.979+0.02 model to predict reddening at early phases despite having a thin helium shell, it was also the best match spectroscopically to SN~2022joj. 

The light curve models do a decent job predicting the luminosity of SN~2022joj in particular Shen 0.9+0.1, although the best spectra model match Shen 0.979+0.02 underpredicts it. Therefore, it is interesting to see that the best light curve model comes from a thick helium shell rather than a thin one like the spectra models. However, part of this discrepancy could be attributed to the uncertainty in the distance to SN~2022joj. From Figure \ref{fig:color_evol}, we see that the color evolution of double detonation models, in particular the models by Ni 1.0+0.01 and Shen 0.971+0.02, predict extreme reddening at early epochs, consistent with SN~2022joj. The reddening does seem to happen a few days earlier than in SN~2022joj, but this can also be attributed to uncertainties in the explosion time of SN~2022joj. Overall, the double detonation models do a good job at predicting the spectra and colors before and at maximum light. After maximum light the colors from the models deviate from data. Again, it is important to note that such deviations are in line with expectations, given the underlying assumption of LTE and incorporate models for completeness.

Mixing during a subsonic explosion can yield iron-group elements in the outer shell of the WD that could explain the reddening observed at the early epochs. This was predicted by some Chandrasekhar-mass explosion models, where a white dwarf initially deflagrates subsonically before transitioning into a detonation \citep{Reinecke02}. Simulations have shown that when deflagrations occur off-center and are asymmetric, they can generate clusters of Fe-peak elements on the surface, which become visible only from certain favorable viewing perspectives \citep{Maeda_2010, Seitenzahl2013}. We compared various $\rm{^{56}Ni}$ distributions to SN~2022joj, but found that shallower $\rm{^{56}Ni}$ predict bluer colors and broader early light curves.  

\subsection{Remote Location}\label{sec:remote_location}
\begin{figure}
\includegraphics[width=0.45\textwidth]{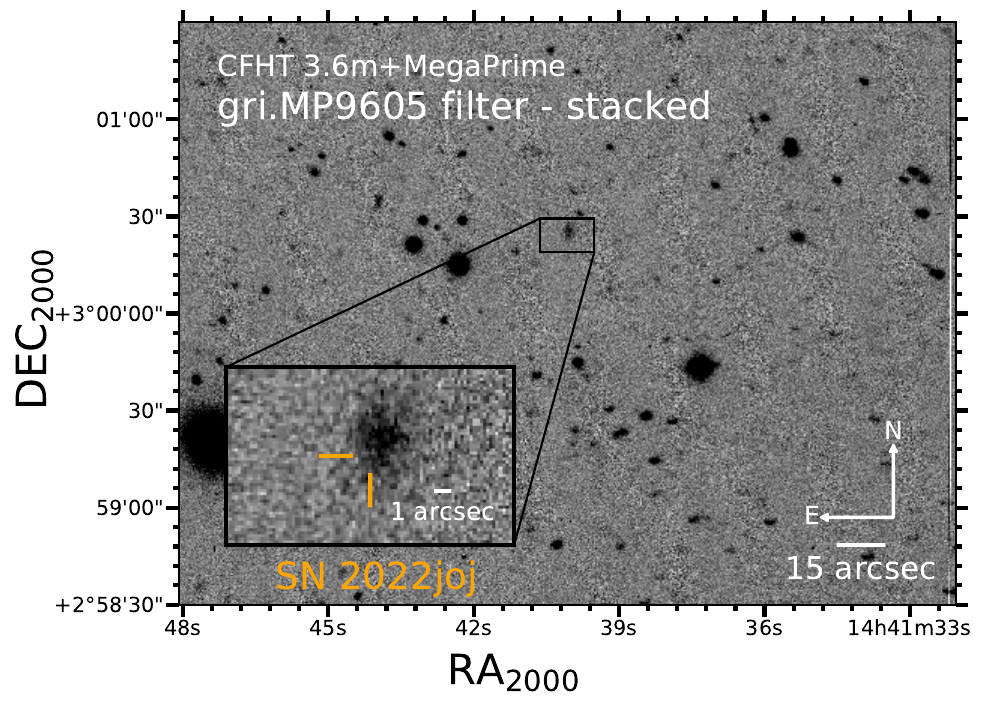}
\caption{The host of SN 2022joj and its surroundings. The image was created by stacking 6 poses from CFHT+MegaPrime, obtained on two different epochs, on 2017, May 19 and on 2021, April 17, both prior SN~2022joj's explosion. The broad \textit{gri} single filter was used for all the images.
\label{fig:host}}
\end{figure}

At first glance, SN~2022joj appears to be hostless with no nearby galaxy as shown in Figure \ref{fig:image}. We were able to retrieve deep images\footnote{The images were obtained from the Canadian Astronomy Data Centre (\url{https://www.cadc-ccda.hia-iha.nrc-cnrc.gc.ca/en/})} secured by the Canada France Hawaii Telescope (CFHT), using MegaPrime and their broad single \textit{gri} filter. The field was observed 6 times over 2 epochs (2017, May 19 and 2021, April 17), and we stacked them in order to achieve a deeper image. The result is shown in Figure \ref{fig:host}, where a spacially resolved host galaxy is clearly detected.
The host appears to be detected in the Pan-STARRS Legacy Survey, which report a Kron magnitude of 20.8 mag in the $i$-band. This apparent magnitude corresponds to an absolute magnitude of -14.3 mag, based on our estimate distance of SN~2022joj (105.2 Mpc), indicating a small dwarf galaxy \citep{Sabatini2003}. 


Interestingly, a few double detonation candidates have been found far from their host galaxies --- this is considerable given that there is only a handful of double detonation candidates in literature. Notable examples include OGLE-2013-SN-079 \citep{Inserra_2015}, SN 2018byg \citep{De_2019}, and SN 2016dsg \citep{Dong2022}. SNe with significant displacements have also been detected in numerous Ca-strong transients. One explanation for these Ca-rich transients (theorized to be single detonations of WDs) is that they originate from high-velocity, kicked systems, and explode at considerable distances from their original location within the host galaxy prior to their occurrences \citep{Lyman2014}. 


Some alternative progenitor system with a double detonation include: a hot subdwarf B (sdB) binary with a WD companion \citep{Geier2013, Kupfer2022}, a WD in a dynamically unstable system where the secondary is a either He WD or a hybrid between He/CO \citep{Guillochon_2010, pakmor2013}, and a potential outcome is a dynamically driven double degenerate double-detonation ($\rm{D^{6}}$) where the companion WD survives explosion and is flung away \citep{shen_d6_2018}, and a WD accreting mass from a He star \citep{Neunteufel2016, Polin2019}. 

\section{Conclusions} \label{sec:conclusion}

We presented photometric and spectroscopic data of SN~2022joj, a peculiar SN Ia. We tested models with shallow nickel distributions and found that they tend to match the brightness of the supernova, but show longer early rise times and brighter early B band flux, contrary to SN~2022joj's behavior. The double detonation models, in particular models with a thin He-shell, were able to explain the light curve properties of SN~2022joj. Notably, the thin He-shell model offers a more accurate explanation for the spectroscopic features and color evolution observed, in contrast to the thick He-shell model. The double detonation scenario is an attractive explosion mechanism, as it is able to explain the peculiarities of these early observations, especially the early reddening as shown in Figure \ref{fig:color_evol}. Nebular and near-infrared spectra also reveal clues for the progenitor of SN~2022joj. However, neither of the spectra disclose whether SN~2022joj truly is a double detonation. In the NIR spectra, it is unclear if there is any unburned He left from the explosion or whether there may be a possible blend between \ion{He}{1} and \ion{Mg}{2}. The nebular spectra, on the other hand, pose more of a mystery. Based on the \cite{Polin_2021} nebular spectra models of double detonations with a thin He-shell, it is expected that the [\ion{Ca}{2}] emission feature should be much stronger than observed in SN~2022joj. Moreover, SN 2018aoz does not show a strong [Ca II] either despite being a double detonation candidate.  Further modeling is required to test if the double detonation models can explain these objects at all phases, especially nebular observations.

\begin{acknowledgments}

We are grateful for the National Science foundation (NSF) and the University of California, Santa Barbara (UCSB) for funding this project through  NSF grant AST-1911225, AST-1911151, and the Central Campus Fellowship.

This research makes use of observations from the Las Cumbres Observatory network, in addition to the MARS ZTF alert broker developed by Las Cumbres Observatory software engineers.

This research has made use of the NASA/IPAC Extragalactic Database (NED) which is operated by the Jet Propulsion Laboratory, California Institute of Technology, under contract with NASA.

This research used the facilities of the Canadian Astronomy Data Centre operated by the National Research Council of Canada with the support of the Canadian Space Agency.

Time domain research by D.J.S, and the University of Arizona team is supported by NSF grants AST-1821987, 1813466, 1908972, and 2108032, and by the Heising-Simons Foundation under grant \#2020--1864.

Financial support for K.J.S. was in part provided by NASA/ESA Hubble Space Telescope program \#15871

Research by Y.D., and S.V.,  and N.M.R is supported by NSF grant AST-2008108.

S.J.B. and D.M.T. acknowledge support from NASA grant HST-AR-16156.

\end{acknowledgments}

\vspace{5mm}
\facilities{HST(STIS), Swift(XRT and UVOT), AAVSO, CTIO:1.3m,
CTIO:1.5m,CXO}

\software{astropy \citep{2013A&A...558A..33A,2018AJ....156..123A},  
          Cloudy \citep{2013RMxAA..49..137F}, 
          Source Extractor \citep{1996A&AS..117..393B},
          {\tt YSE-PZ} \citep{CoulterZenodo, CoulterYSEPZ},
          }

\appendix

\renewcommand\thetable{A\arabic{table}} 
\setcounter{table}{0}

\begin{table*}
\begin{center}
\begin{tabular}{ccccccc}
\hline
\hline
JD & epoch & \textit{B} & \textit{V} & \textit{g} & \textit{r} & \textit{i} \\
\hline
2459711 & -11 & 18.62 (0.05) & 17.52 (0.04) & 17.92 (0.03) & 17.58 (0.03) & 18.0 (0.03)  \\
2459712 & -10 & 17.90 (0.03) & 17.16 (0.02) & 17.35 (0.01) & 17.34 (0.02) & 17.76 (0.04) \\
2459713 &  -9 & 17.37 (0.04) & 16.94 (0.04) & 16.97 (0.03) & 17.13 (0.03) & 17.44 (0.05) \\
2459717 &  -5 & 16.42 (0.01) & 16.35 (0.01) & 16.19 (0.01) & 16.49 (0.01) & 16.8 (0.01)  \\
2459718 &  -4 & 16.28 (0.01) & 16.23 (0.02) & 16.06 (0.01) & 16.36 (0.01) & 16.68 (0.01) \\
2459720 &  -2 & 16.12 (0.01) & 16.08 (0.02) & 15.91 (0.01) & 16.19 (0.01) & 16.56 (0.01) \\
2459723 &   1 & 16.02 (0.02) & 15.96 (0.02) & 15.81 (0.01) & 16.06 (0.01) & 16.6 (0.02)  \\
2459725 &   3 & 16.07 (0.02) & 15.95 (0.02) & 15.85 (0.01) & 16.03 (0.01) & 16.65 (0.02) \\
2459730 &   8 & 16.47 (0.02) & 16.1 (0.02)  & 16.11 (0.01) & 16.15 (0.01) & 16.86 (0.03) \\
2459733 &  11 & 16.72 (0.02) & 16.21 (0.02) & 16.35 (0.01) & 16.27 (0.01) & 17.02 (0.02) \\
2459735 &  12 & 16.95 (0.03) &     $-$      &     $-$      &     $-$      &     $-$      \\
2459736 &  14 & 17.11 (0.02) & 16.46 (0.02) & 16.64 (0.01) & 16.53 (0.01) & 17.28 (0.02) \\
2459738 &  16 & 17.44 (0.02) & 16.65 (0.02) & 16.89 (0.01) & 16.68 (0.01) &     $-$      \\
2459739 &  16 &     $-$      &     $-$      &     $-$      &     $-$      & 17.33 (0.02) \\
2459740 &  17 & 17.46 (0.03) & 16.66 (0.02) & 16.97 (0.01) & 16.69 (0.01) & 17.26 (0.03) \\
2459741 &  18 & 17.69 (0.05) & 16.81 (0.03) & 17.21 (0.02) & 16.85 (0.03) & 17.36 (0.05) \\
2459743 &  20 & 17.85 (0.04) & 16.91 (0.04) & 17.4 (0.03)  & 16.83 (0.02) & 17.39 (0.04) \\
2459744 &  21 & 17.85 (0.06) & 16.92 (0.04) & 17.39 (0.03) & 16.85 (0.03) & 17.36 (0.06) \\
2459746 &  23 & 18.09 (0.07) & 17.06 (0.06) & 17.62 (0.04) & 16.94 (0.04) & 17.48 (0.12) \\
2459748 &  25 & 18.23 (0.03) & 17.16 (0.02) & 17.7 (0.01)  & 17.06 (0.07) & 17.26 (0.04) \\
2459753 &  31 & 18.63 (0.03) & 17.49 (0.02) & 18.13 (0.01) & 17.23 (0.01) & 17.47 (0.02) \\
2459757 &  35 & 18.81 (0.05) & 17.71 (0.03) & 18.35 (0.03) & 17.4 (0.02)  & 17.57 (0.03) \\
2459767 &  44 &     $-$      & 17.98 (0.17) &     $-$      & 17.88 (0.09) &     $-$      \\
2459772 &  50 & 19.41 (0.09) & 18.3 (0.05)  & 18.8 (0.05)  & 18.23 (0.05) & 18.56 (0.07) \\
2459778 &  56 & 19.4 (0.04)  & 18.48 (0.04) & 18.91 (0.03) & 18.41 (0.03) & 18.78 (0.06) \\
2459784 &  62 & 19.62 (0.07) & 18.68 (0.03) & 19.04 (0.03) & 18.56 (0.03) & 18.9 (0.08)  \\
2459787 &  65 & 19.63 (0.03) & 18.74 (0.04) & 19.08 (0.02) & 18.63 (0.02) & 18.99 (0.11) \\
2459796 &  73 & 19.82 (0.05) & 19.07 (0.04) & 19.19 (0.03) & 18.99 (0.03) & 19.35 (0.05) \\
2459801 &  78 &     $-$      & 19.17 (0.11) & 19.3 (0.08)  & 19.08 (0.08) &     $-$      \\
2459805 &  83 & 19.91 (0.04) &     $-$      &     $-$      &     $-$      &     $-$      \\
2459806 &  83 &     $-$      & 19.22 (0.04) & 19.36 (0.03) & 19.22 (0.05) & 19.78 (0.1)  \\
2459810 &  88 & 19.97 (0.04) & 19.31 (0.04) & 19.45 (0.02) & 19.39 (0.03) &     $-$      \\
2459811 &  88 &     $-$      &     $-$      &     $-$      & 19.42 (0.04) & 19.84 (0.07) \\

\hline
\end{tabular}
\caption{Phase with respect to the B max where the dates are rounded up.}
\label{tab:data}
\end{center}
\end{table*}

\begin{table*}
\begin{center}
\begin{tabular}{cccccccc}
\hline
\hline
JD & epoch & \textit{UV-W2} & \textit{UV-M2} & \textit{UV-W1} & \textit{U} & \textit{B} & \textit{V} \\
\hline
2459716.88 & -5.42  & $>20.68$ & $>20.81$ & $19.19\,(0.29)$ & $17.26\,(0.09)$ & $16.37\,(0.04)$ & $16.38\,(0.06)$\\
2459720.56 & -1.74  & $19.37\,(0.15)$ & $>20.37$ & $18.14\,(0.11)$ & $16.55\,(0.06)$ & $16.05\,(0.03)$ & $16.03\,(0.04)$\\
2459720.96 & -1.34  & $19.43\,(0.15)$ & $19.59\,(0.22)$ & $18.19\,(0.11)$ & $16.70\,(0.06)$ & $16.03\,(0.03)$ & $15.93\,(0.04)$\\
2459723.25 & 0.95  & $19.35\,(0.13)$ & $20.22\,(0.30)$ & $18.52\,(0.16)$ & $16.57\,(0.05)$ & $15.99\,(0.03)$ & $15.94\,(0.03)$\\
\hline
\end{tabular}
\caption{\textit{Swift}+UVOT photometry. Magnitudes are in Vega system. Phase are with respect to the B max.}
\label{tab:data}
\end{center}
\end{table*}

\clearpage
\bibliography{sample631}{}
\bibliographystyle{aasjournal}

\end{document}